\documentclass[seceq]{ptptex}

\usepackage{graphicx}

\title{
Chiral Thermodynamics of Dense QCD%
}

\author{
Chihiro \textsc{Sasaki}%
}

\inst{
Frankfurt Institute for Advanced Studies,
D-60438 Frankfurt am Main,
Germany
}

\abst{
We show a phase diagram of a two-flavored parity doublet model 
on top of the nuclear matter ground state within the mean field approximation.
An exotic phase with unbroken center symmetry of chiral group is also 
discussed. When crossing this phase boundary the baryon number susceptibility
exhibits a characteristic feature.
}

\begin{document}

\maketitle

\section{Parity doubled nucleons}
\label{sec:int}

Model studies of hot and dense matter have suggested a rich phase structure 
of QCD at temperatures and quark chemical potentials of order 
$\Lambda_{\rm QCD}$. Our knowledge on the phase structure however remains 
limited and the description of strongly interacting matter does not 
reach a consensus yet~\cite{qmproc}. 
In particular, properties of baryons near the chiral symmetry restoration are
poorly understood.
The realistic modeling of dense baryonic matter must take into
account the existence of the nuclear matter saturation point, i.e. the bound
state at baryon density $\rho_0 = 0.16$ fm$^{-3}$, like in Walecka type
models~\cite{walecka}.
Several chiral models with pure hadronic degrees of freedom~\cite{other,nnjl}
have been constructed in such a way that the nuclear matter has the true 
ground state. An alternative approach is to describe a nucleon as
a dynamical bound-state of a diquark and a quark~\cite{bentz}.

In the mirror assignment of chirality to nucleons~\cite{dk,mirror},
dynamical chiral symmetry breaking generates a mass difference between parity 
partners and the chiral symmetry restoration does not necessarily dictate the 
chiral partners to be massless. Mirror baryons embedded in linear and non-linear 
chiral Lagrangians have been applied to study their phenomenology 
in vacuum~\cite{dk,mirror,lsma1}, nuclear matter~\cite{hp,pdm} 
and neutron starts~\cite{astro}.
Identifying the true parity partner of 
a nucleon is also an issue. In the mirror models $N(1535)$ is usually taken to 
be the negative parity state. This choice however fails to reproduce the decay 
width to a nucleon and $\eta$. This might indicate another negative parity 
state lighter than the $N(1535)$~\cite{pdm}, which has not been observed so far.

\begin{figure}
\begin{center}
\includegraphics[width=6cm]{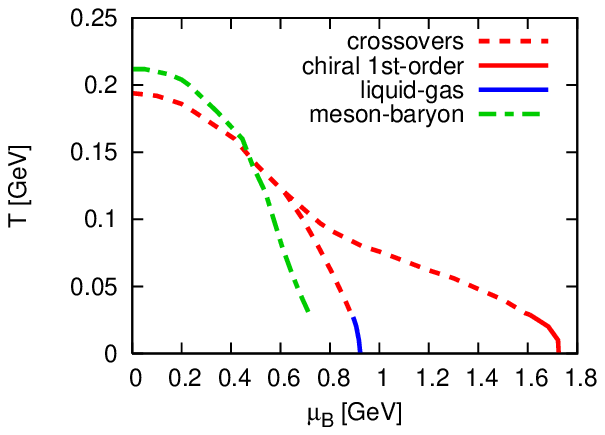}
\includegraphics[width=6cm]{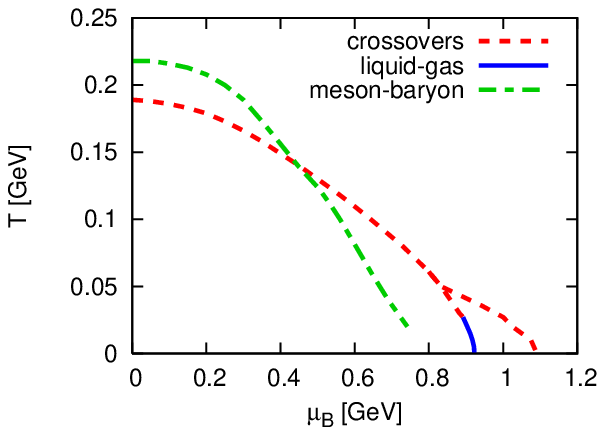}
\caption{
The phase diagram in the parity doublet model~\cite{pdmour}.
The mass of the negative parity nucleon was taken to be $1.5$ GeV (left)
and $1.2$ GeV (right).
}
\label{pdm}
\end{center}
\end{figure}

The parity doublet model has been applied to a hot and dense hadronic 
matter and the phase structure of a chiral symmetry restoration as well as 
a liquid-gas transition of nuclear matter was explored~\cite{pdmour}. 
In Fig.~\ref{pdm} we show the phase diagram for two different masses of
the negative parity state, $m_{N-}=1.5$ GeV and $1.2$ GeV. The latter
is considered to be an phenomenological option.
At zero temperature the system experiences a first-order liquid-gas
transition at $\mu_B = 923$ MeV and the baryon density shows a jump from
zero to a finite value $\rho \neq 0$. The order of chiral phase transition and its
location depend on the set of parameters, especially on mass of the negative 
parity state. 
Roughly speaking this phase transition occurs when the baryon chemical potential
reaches the mass of the negative parity state, $\mu_B \sim m_{N-}$.
If we take the most frequently used value
$m_{N-}=1500$ MeV, then in addition to the nuclear liquid-gas phase transition
we obtain a weak first-order chiral transition 
at $\rho \sim 10\,\rho_0$. 
With a lower mass $m_{N-}=1200$ MeV 
we get no true chiral phase transition but only a crossover at much lower density
$\rho \sim 3\,\rho_0$.
The liquid-gas transition survives up to $T = 27$ MeV. Above this temperature
there is no sharp phase transition but the order parameter is still attracted 
by the critical point:
the order parameter typically shows a double-step structure and
this makes an additional crossover line terminating at the liquid-gas 
critical point. Another crossover line corresponding to the chiral symmetry 
restoration follows 
the steepest descent of the second reduction in $\langle \sigma \rangle$. 
With increasing
temperature the two crossover lines become closer and finally merge.

In contrast, the trajectory of a meson-to-baryon ``transition'' defined 
from the ratio of particle number densities is basically driven by
the density effect with the hadron masses being not far from their vacuum
values. The line is almost independent of the parameter set and goes rather
close to the liquid-gas transition line.
The chiral crossover and the meson-baryon transition lines intersect
at $(T,\mu_B) \sim (150,450)$ MeV. 
The parity doublet model thus describes 3 domains: a chirally broken
phase with mesons thermodynamically dominating, another chirally broken phase 
where baryons are more dominant and the chirally restored phase,
which can be identified with quarkyonic matter~\cite{quarkyonic}.
It is worthy to note that this intersection point is fairly
close to the estimated triple point at which hadronic matter, quarkyonic matter
and quark-gluon plasma may coexist~\cite{triple}.

\section{Role of the tetra-quark at finite density}
\label{sec:znf}

There is a possibility of two different phases {\it with broken
chiral symmetry} at finite density where a rather unorthodox pattern 
of spontaneous chiral symmetry breaking could be realized~\cite{shifman}.
This pattern keeps the center of chiral group unbroken, i.e.
\begin{equation}
SU(N_f)_L \times SU(N_f)_R \to SU(N_f)_V \times (Z_{N_f})_A\,,
\label{breaking}
\end{equation}
where a discrete symmetry $(Z_{N_f})_A$ is the maximal axial
subgroup of $SU(N_f)_L \times SU(N_f)_R$. The center $Z_{N_f}$ 
symmetry protects a theory from condensate of quark bilinears 
$\langle \bar{q}q \rangle$. Spontaneous symmetry breaking
is driven by quartic condensates which are invariant under
both $SU(N_f)_V$ and $Z_{N_f}$ transformation.
In a system with the breaking pattern~(\ref{breaking})
the quartic condensate is the strict order parameter which
separates different chirally-broken phases~\footnote{
 A similar phase structure was found in an O(2) model~\cite{watanabe}
 and in the Skyrme model on crystal~\cite{skyrmion}.
}.

\begin{figure}
\begin{center}
\includegraphics[width=6cm]{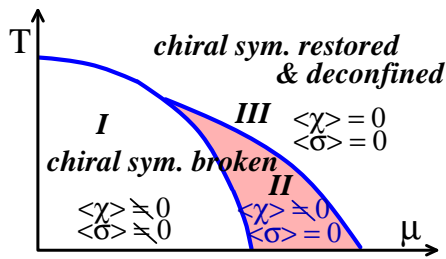}
\caption{
Schematic phase diagram mapped onto $(T,\mu)$ plane.
}
\label{phase}
\end{center}
\end{figure}

Assuming~(\ref{breaking})  
at finite density,
it has been shown that an intermediate phase between
chiral symmetry broken and its restored phases can be realized
using a general Ginzburg-Landau free energy~\cite{hst}.
The pion decay constant is read from the Noether current as
$F_\pi = \sqrt{\sigma_0^2 + \frac{8}{3}\chi_0^2}$
with $\chi_0$ and $\sigma_0$ being the expectation values
of 4-quark and 2-quark scalar fields, determined from the gap equations.
The effective potential deduced in the mean field approximation
describes three distinct phases characterized by the two order
parameters: Phase I represents the system where both chiral 
symmetry and its center are spontaneously broken due to 
non-vanishing expectation values $\chi_0$ and $\sigma_0$. 
The center symmetry is restored when $\sigma_0$ becomes zero. 
However, chiral symmetry remains broken as long as $\chi_0$ is non-vanishing,
where the pure 4-quark state is the massless
Nambu-Goldstone boson (phase II). The chiral symmetry restoration takes
place under $\chi_0 \to 0$ which corresponds to phase III.
With an explicit breaking of chiral symmetry
one would draw a phase diagram mapped onto $(T,\mu)$ plane
as in Fig.~\ref{phase}.
A characteristic feature with unbroken center symmetry is found in
the baryon number susceptibility. The $Z_2$ invariance prohibits
a Yukawa coupling of $\chi$ to baryons.
Consequently, the baryon number susceptibility exhibits a maximum
when across the $Z_2$ cross over.

\section{Summary and Discussions}
\label{sec:sum}

The parity doublet model within the mean field approximation describes 
the nuclear matter ground state at zero temperature and a chiral crossover 
at zero chemical potential at a reasonable temperature, which are the minimal 
requirements to describe the QCD thermodynamics. The first-order phase 
transitions appear only at low temperatures, below $T \sim 30$ MeV. 
Nevertheless, at higher temperature they still affect the order parameter 
which exhibits a substantial decrease near the liquid-gas {\it and} chiral 
transitions. If the chiral symmetry restoration is of first order, 
criticality around the end points of the two first-order phase transitions
will be the same due to the identical universality class~\cite{cp}.

A possibility of a non-standard breaking pattern leads 
to a new phase where chiral symmetry is spontaneously broken 
while its center symmetry is restored. This might appear as an 
intermediate phase between chirally broken and restored phases 
in $(T,\mu)$ plane. The appearance of this phase also suggests 
a new critical point in low temperatures~\cite{hst}.
A tendency of the center symmetry restoration is carried by
the net baryon number density which shows a rapid increase
indicating baryons more activated,
and this is reminiscent of the quarkyonic transition.

\section*{Acknowledgments}
I am grateful for fruitful collaboration with 
M.~Harada, I.~Mishustin and S.~Takemoto.
Partial support by the Hessian
LOEWE initiative through the Helmholtz International
Center for FAIR (HIC for FAIR) is acknowledged.

%

\end{document}